# Sub-Terahertz Spin-Pumping from an Insulating Antiferromagnet

Priyanka Vaidya,[1] Sophie A. Morley,[2] Johan van Tol,[3] Yan Liu,[4] Ran Cheng,[5] Arne Brataas,[6] David Lederman,[2] and Enrique del Barco,[1*]

[1]*Department of Physics, University of Central Florida, Orlando, FL 32765, USA.*

[2]*Department of Physics, University of California Santa Cruz, Santa Cruz, CA 95064, USA.*

[3]*National High Magnetic Field Laboratory, Florida State University, Tallahassee, FL 32310, USA.*

[4]*College of Sciences, Northeastern University, Shenyang, Liaoning, China*

[5]*Department of Electrical and Computer Engineering, and Department of Physics and Astronomy, University of California, Riverside, CA 92521, USA.*

[6]*Department of Physics, Center for Quantum Spintronics, Norwegian University of Science and Technology, NO-7491 Trondheim, Norway*

*To whom correspondence should be addressed: E-mail: delbarco@physics.ucf.edu*

**Abstract:**

Spin-transfer torque and spin Hall effects combined with their reciprocal phenomena, spin-pumping and inverse spin Hall (ISHE) effects, enable the reading and control of magnetic moments in spintronics. The direct observation of these effects remains elusive in antiferromagnetic-based devices. We report sub-terahertz spin-pumping at the interface of a uniaxial insulating antiferromagnet $MnF_2$ and platinum. The measured ISHE voltage arising from spin-charge conversion in the platinum layer depends on the chirality of the dynamical modes of the antiferromagnet, which is selectively excited and modulated by the handedness of the circularly polarized sub-THz irradiation. Our results open the door to the controlled generation of coherent pure spin currents at THz frequencies.

In the absence of external magnetic fields and below their Néel ordering temperatures, antiferromagnetic (AF) systems exhibit magnetic order with zero net magnetization *(1)*. Unlike ferromagnets, AF materials do not produce stray magnetic fields, and therefore AF elements can be tightly packed while operating independently without crosstalk. AF elements also have a low magnetic susceptibility and thus are immune against external magnetic perturbations. Another salient advantage of AF materials when compared to ferromagnetic systems is that in ferromagnets spin dynamics is governed by external, dipolar, and anisotropy fields (typically limited to GHz frequencies), whereas in AF materials spin dynamics depends on the combined effect of magnetic anisotropy and the substantial exchange interaction, which leads to spin excitations in the much higher THz frequency range. This "exchange amplification" phenomenon allows for the control of ultrafast AF dynamics with moderate external currents *(2, 3)*, making antiferromagnets an appealing choice for the generation, detection, and modulation of coherent THz signals *(4-6)*. Historically, the THz region of the electromagnetic spectrum has been difficult to exploit *(7)*.

For decades, AF systems have been an auxiliary element in spintronic devices such as the passive exchange bias layer in spin valves *(8)*. Although a few reports of AF anisotropic magnetoresistance showed that AF materials could indeed be employed to store magnetic information *(3, 9-12)*, it remains an open question whether they can be utilized as active ingredients directly controllable through electrical currents. Recently, there has been progress in this direction owing to the experimental realizations *(11, 13, 14)* of spin-orbit torques in AF systems with special lattice symmetries *(15)*. Moreover, it has been demonstrated that insulating AF hematite ($\alpha$-$Fe_2O_3$) supports unprecedented long-range spin transport across micrometers *(16)*. Nevertheless, a crucial missing piece is the experimental observation of spin-transfer torque *(17, 18)* and its reciprocal effect— dynamical spin-pumping *(19)*, which are fundamental to manipulate the AF order parameter by electrical means. Ross *et al*. reported experiments in an AF ($MnF_2$/Pt) heterostructure showing a small difference of the inverse spin Hall effect signals upon reversal of the magnetic field, which was consistent with but not unique to coherent spin-pumping *(20)*.

Here we demonstrate sub-THz dynamical generation and injection of pure spin currents—coherent spin-pumping—from a crystalline $MnF_2$ AF insulator layer into a heavy metal platinum thin film, where strong spin-orbit coupling enables spin-charge current interconversion through the inverse spin Hall effect (ISHE).

**Antiferromagnetic Resonance in Insulating Antiferromagnet $MnF_2$**

Below the Néel temperature $T_N$, the long-range magnetic order in simple collinear AF systems, like uniaxial insulating $MnF_2$, results from the exchange interaction that favors anti-parallel alignment between neighboring sublattice magnetizations ($\vec{M}_1$ and $\vec{M}_2$). In contrast to ferromagnets, the total magnetization $\vec{M} = \vec{M}_1 + \vec{M}_2$ vanishes, and the AF order parameter is represented by $\vec{L} = \vec{M}_1 - \vec{M}_2 \neq 0$, known as the Néel vector *(1)*. According to the theories of Keffer and Kittel *(21)*, and Nagamiya *et al*. *(22)*, the resonance frequencies of the uniform precessional modes (with wave vector $k = 0$) are:

$$\omega_{res} = \gamma\mu_0\sqrt{H_A(2H_E + H_A)} \pm \gamma\mu_0 H \quad \text{for } H < H_{SF} \quad (1)$$
$$\omega_{res} = \gamma\mu_0\sqrt{H^2 - 2H_E H_A} \quad \text{for } H > H_{SF} \quad (2)$$

where $\gamma$ is the gyromagnetic ratio, $H$ is the external magnetic field applied along the easy (anisotropy) axis, $H_A$ and $H_E$ are the effective fields associated with the uniaxial anisotropy and the AF exchange interaction, respectively, and $\mu_0$ is the permeability of free space. The so-called spin-flop field, $H_{SF} = \sqrt{2H_E H_A}$, separates the AF dynamics into two distinct regimes. For $H < H_{SF}$, Eq. 1 describes the frequencies of the two dynamical modes exhibiting opposite chiralities, which are split by a longitudinal magnetic field $H$ into a high-frequency mode and a low-frequency mode. As illustrated in the upper-right inset of Fig. 1, the high- (low-) frequency mode has the right- (left-) handed chirality with respect to the magnetic field, a large precessional cone angle for $\vec{M}_1$ ($\vec{M}_2$), and a spin-down (-up) angular momentum. The frequency separation of the two modes increases linearly with an increasing applied field until a sufficiently strong field ($H = H_{SF}$) renders the ground state configuration unstable. At this critical point, the sublattice magnetic moments abruptly flop towards the normal plane of $H$ and both are canted towards $H$, so that $\vec{L}$ becomes perpendicular to the anisotropy axis. If the applied field is not perfectly aligned with the easy axis, the spin-flop transition broadens into a finite window in which the HFM drops rapidly with an increasing $H$ (as can be observed in Fig. 1). The resonance frequency in the $H > H_{SF}$ regime (often referred to as the quasi-ferromagnetic mode (QFM)) grows with an increasing $H$ as shown in Eq. 2. In accordance with the above picture, we categorize the dynamical modes in an AF system into three characteristic regimes: *i*) the low- and high-frequency modes (LFM and HFM) with opposite chiralities for $H < H_{SF}$; *ii*) the spin-flop (SF) mode residing in the narrow window of the spin-flop transition; and, *iii*) the QFM mode for $H > H_{SF}$ (after the SF transition is completed *(23)*).

We computed the magnetization dynamics for each magnetic sublattice of a uniaxial antiferromagnet MnF$_2$ by solving the Landau-Lifshitz equation:

$$\frac{d\vec{M}_i}{dt} = -\mu_o \gamma \vec{M}_i \times \vec{H}_{eff} - \alpha \frac{\mu_o \gamma}{M_s} \left( \vec{M}_i \times \left( \vec{M}_i \times \vec{H}_{eff} \right) \right) \quad (3)$$

with an effective field $\vec{H}_{eff}$ comprising the exchange field ($H_E = 47.05$ T), the anisotropy field ($H_A = 0.82$ T), the externally applied field ($H$), and the microwave field ($\vec{H}_m = (H_o \cos(2\pi ft), H_o \sin(2\pi ft + \theta), 0)$ ), where the polarization is determined by changing the phase factor $\theta$ from 0 to $2\pi$, and $i =$1, 2 labels the two-sublattices. We used the following parameter values for the calculation: $\gamma = \gamma_e$, saturation magnetization $M_s = 47.7$ kA/m, and $\alpha = 0.001$, in agreement with previously reported values *(24, 25)*. The theoretical results are displayed in Fig. 1 together with the measured spectroscopic antiferromagnetic resonance (AFMR) absorptions; the experimental data are represented by solid symbols corresponding to three different samples studied at four available frequencies (horizontal orange arrows). Figure S1 shows the corresponding spectra. The upper-left inset to Fig. 1 shows the electron paramagnetic resonance (EPR) spectrum obtained at $f = 395$ GHz (red curve) for the magnetic field range corresponding to the HFM resonance (blue triangle at $\mu_0 H = 4.70$ T). The EPR signal is significantly distorted by saturation of the probe thanks to the large thickness of the MnF$_2$ single crystal used in these experiments but still allows us to determine the location of the resonances spectroscopically *(26)*. The results agree well with the theoretical calculations and are in excellent agreement with previously published antiferromagnetic resonance data *(25, 27)* and theoretical analyses *(1, 22)* reported for MnF$_2$.

**Coherent Spin-Pumping and the Inverse Spin Hall Effect in MnF$_2$/Pt**

Coherent spin-pumping *(28-31)* has been central to the advance of ferromagnetic-based spintronics; it serves as a tool to generate spin currents dynamically, avoiding, for example, conductance mismatch issues at the interface between magnetic and non-magnetic materials. In the realm of antiferromagnetic-based spintronics, Cheng *et al*. developed a theoretical framework to understand dynamical spin injection from an AF material undergoing coherent precession (AFMR) into an adjacent non-magnetic material *(19)* (see also Ref.(*32*)). Contrary to the conventional wisdom that spin-pumping from antiparallel sublattice spins would cancel out, ref. (19) established that coherent resonant rotations of different sublattice spins contribute constructively to the pumped spin current. A heuristic understanding of AF spin-pumping is that spin currents pumped from the two sublattice magnetization are proportional to $\vec{M}_1 \times \dot{\vec{M}}_1$ and $\vec{M}_2 \times \dot{\vec{M}}_2$, respectively, if we view $\vec{M}_1$ and $\vec{M}_2$ as two independent ferromagnets. As illustrated in Fig. 1 (upper-right inset), the two sublattice rotate in the same angular direction with a $180^0$ phase difference, thus $\vec{M}_1 \approx -\vec{M}_2$ and $\dot{\vec{M}}_1 \approx -\dot{\vec{M}}_2$. Consequently, contributions from the two sublattices add up, yielding the total pumped spin current proportional to $\vec{L} \times \dot{\vec{L}} + \vec{M} \times \dot{\vec{M}}$. Because $H_E \gg H_A$ in MnF$_2$, we have $|\vec{L}| \gg |\vec{M}|$ and $\vec{M}$ can be approximately expressed in terms of $\vec{L}$ as $\vec{M} \approx \left[\frac{H}{H_E}\vec{L} \times (\hat{z} \times \vec{L}) - \frac{1}{\gamma\mu_0 H_E}\vec{L} \times \dot{\vec{L}}\right]$, from which one can tell that $\vec{L} \times \dot{\vec{L}}$ is much larger than $\vec{M} \times \dot{\vec{M}}$. That is to say, it is the Néel vector $\vec{L}$, rather than the vanishingly small magnetization $\vec{M}$, that generates the most essential part of coherent spin-pumping. Furthermore, it was predicted in Ref. (*19*), that the polarization of the driving ac field determines the direction of the pumped spin current. Dynamical modes with opposite chirality coexist in a collinear AF system at zero field; and they can be selectively excited by an ac field with matching polarization. In other words, spins are pumped with opposite polarizations depending on whether the right- or left-handed mode is excited (by a right- or left-handed circularly polarized stimulus). A magnetic field breaks the degeneracy between the opposite chirality modes. Consequently, only the correct combination of the irradiation frequency and handedness excites a particular AF mode. Therefore, depending on the handedness of the circular polarization and the frequency of irradiation at a given magnetic field, opposite spin currents would be generated in the adjacent non-magnetic material and transform into opposite ISHE electric signals.

In the following, we discuss the measurements of the electrical signals observed by sweeping the magnetic field while irradiating $MnF_2$/Pt samples with circularly polarized sub-THz microwaves of frequency $f$. The measured ISHE spectra in samples 3 and 2 are shown in Figs. 2, A and B, ($f = 395$ GHz) and Figs. 2, C and D, ($f = 240$ GHz), respectively. Figure S2 shows the power dependence data for $f = 395$ GHz. For $f = 240$ GHz, clear voltage signals were observed associated with the spectra for the LFM, the SF mode, and the QFM. All signals reversed sign when the applied magnetic field reversed direction, which is consistent with the time reversal symmetry. However, the signal magnitudes differed for opposite handedness of the microwave stimuli, suggesting that chiral AF modes were selectively excited according to the circular polarization. This contrasting magnitude becomes striking in Figs. 2, C and D, where the LFM only appears at a positive (negative) field ($\mu_0|H| = 0.80$ T) for the left (right)-handed irradiation. This is indeed the expected behavior of a circularly-polarized AF mode in the presence of an external magnetic field. For positive (negative) fields, the LFM mode's chirality is left (right)-handed as it has a spin angular momentum parallel to the magnetic field, whereas the opposite is true for the HFM. There is also a noticeable difference in the strength of the SF signals by reversing only the magnetic field or only the circular polarization. On the other hand, the magnitude of the QFM resonance remains nearly constant, which we will discuss further below.

**Coherent Spin-Pumping vs Incoherent Spin-Seebeck Effect**

A central question arises from these observations: Do the voltage signals originate from coherent spin-pumping at the $MnF_2$/Pt interface, or the incoherent spin Seebeck effect *(33, 34)* induced by a temperature gradient resulting from microwave heating? In ferromagnets, this is quite a challenging question because only the right-handed mode exists,; therefore both coherent and incoherent contributions have the same spin polarization that electrical measurements alone cannot distinguish *(35)*. In this setting, one would need to perform control experiments, such as changing the layers stacking order or conducting thermal transport measurements. The situation is fundamentally different in antiferromagnets. The coexistence of both chiral modes in AF systems allows us to discern between coherent and incoherent contributions from the electrical measurements alone. The high frequency (395GHz) data for sample 3 in Fig. 2, A and B (see analogous data for sample 2 in Fig. S1 and related discussion) indicate that electric signals from the HFM and LFM resonances (at $H = \pm 4.7$ T) behave in exact opposite ways when switching handedness. The HFM signal appears only at positive (negative) fields with the right (left)-handed irradiation *(36)*, as it corresponds to the right (left)-handed chirality of the excited AF mode. In contrast, the sign of spin Seebeck effect would be independent of the microwave handedness, because it primarily originates from the LFM mode (thermally more populated than the HFM) even if the microwave heating stems from the resonant absorption of microwave energy by the HFM. Because different frequencies need to be applied to excite the LFM and HFM, and a magnetic field is present, it is possible that the absorption of electromagnetic energy differs for each mode/polarization causing different heating patterns (*i.e.*, different spin Seebeck signals). Although this could account for the observed modulation of the ISHE signals of the two modes, such incoherent thermal effect cannot account for the complete reversal of the ISHE sign. Therefore, our experimental observation demonstrates that the effect originates from coherent spin-pumping and the ISHE in Pt.

Given the coherent origin of the signals, we can further estimate the spin-mixing conductance of the $MnF_2$/Pt interface from the measured ISHE voltage. Taking into account the back-flow of spin current in the Pt layer (with the spin-mixing conductance $g_r$ in unit of $e^2/h$ per area), we obtain *(4, 19)*:

$$V_{ISHE} = L\theta_S \left(\frac{H_A}{H_E}\right)\left(\frac{\lambda}{d_N}\right)\frac{\hbar e(\gamma B_\perp)^2}{\alpha^2 \omega_R}\frac{g_r \tanh\frac{d_N}{2\lambda}}{\hbar\sigma + 2\lambda e^2 g_r \coth\frac{d_N}{2\lambda}}, \qquad (4)$$

where $L$ is the distance between the two voltage leads; $d_N, \theta_s, \lambda, \sigma$ are the thickness, the spin Hall angle, the spin diffusion length, and the conductivity of the Pt layer; $\alpha$ is the Gilbert damping in MnF$_2$, $\omega_R$ is the angular frequency of AFMR, and $B_\perp$ is the amplitude of magnetic field of the circularly-polarized microwave. Even though it is difficult to acquire the exact value of every parameter, we can estimate the amplitude of $g_r$ with available data reported in the literature. In the MnF$_2$, we have $\frac{H_A}{H_E} \approx 1.8\%$ and $\alpha \approx 0.5 \times 10^{-3}$; in the Pt layer, $d_N \approx 4$ nm, $\theta_s \approx 0.08$, $\lambda \approx 1.4$ nm and $\sigma \approx 4 \times 10^6$ S/m *(37, 38)*. On the peak (dip) point of the 240 GHz resonance, $B_\perp \approx 200$ mG and $V_{ISHE} \approx 25$ nV. Therefore, we obtain $g_r \approx 2.86 \times 10^{18}$ m$^{-2}$, which converts into $\approx 0.66\, e^2/h$ per unit cell area on the interface. This value, though a rough estimate due to uncertainty in some of the parameters, is consistent with the theoretical prediction ($\approx 1\, e^2/h$) *(19, 32)*. Here we point out that the extracted spin-mixing conductance is of a similar magnitude compared to that in ferromagnet/normal metal heterostructures, which confirms the theoretical picture that opposite sublattice magnetizations can constructively pump spins, not cancel.

**Spin-flop mode and the high magnetic field quasi-ferromagnetic mode**

Now we consider the behavior of the SF and QFM signals. Although the handedness of the microwave polarization modulates the SF resonances, it does not affect the QFM signal much. The strength of the intermediate SF resonances, as shown in Fig.

2, is more pronounced when the polarization is right (left)-handed for positive (negative) fields, which is the case for both frequencies. To highlight this result, we show in Fig. 3 all ISHE signals as functions of the relative phase that determines the circular polarization of the microwaves. Whereas in the low-field regime ($H < H_{SF}$) both the LFM and the HFM exhibit oscillatory patterns as a function of the polarization phase (see, e.g., LFM feature in Fig. 3K), in the high-field ($H > H_{SF}$) the QFM signal is essentially constant (e.g., QFM feature on Fig. 3L). On the other hand, the SF signals display a mixture of both regimes—they oscillate on top of a constant background signal that has a similar magnitude as the 240 GHz QFM (see the red arrow in Fig. 3L). The appearance of phase-modulation in the SF resonances is not surprising as they partially retain the features of the HFM (especially the chirality) whereas the ground state undergoes a gradual evolution from the collinear configuration into the spin-flop configuration. However, the sign of the SF resonance of 395 GHz is opposite to that of 240 GHz, which requires a more detailed analysis.

In Fig. 4A, we directly compare the phase-modulation pattern for four particular resonances under positive magnetic fields; Fig. 4B relates them to four representative points on the upper-frequency branch and illustrates their physical meaning. In the low-field regime, as described by point (1), the non-equilibrium (dynamical) spin angular momentum $m_d$ carried by the HFM opposes the magnetic field. In the high-field regime, once the sublattice magnetizations have flopped into a direction perpendicular to the applied magnetic field, as depicted by point (4), a finite (static) magnetization $m_s$ along the magnetic field is induced in the ground state because the Zeeman interaction cants both sublattices towards the applied field. The QFM refers to a right-handed rotation of the induced magnetization similar to a ferromagnet *(21)*, which is why it is named QFM; the sublattice magnetizations are still strongly antiferromagnetically connected by the predominant exchange interaction. Correspondingly, the non-equilibrium spin angular momentum $m_d$ induced upon excitation is negative with respect to the magnetic field; its sign follows that of the HFM. However, the measured ISHE signal arising from the QFM does not follow this rule, indicating that the spin current may not be originating from coherent spin-pumping at point (4). As a theoretical check, we numerically calculated the dc coherent spin-pumping for all points given by the following expression *(19)* :

$$\frac{e}{\hbar}\vec{I}_s = G_r(\vec{n} \times \dot{\vec{n}} + \vec{m} \times \dot{\vec{m}}) - G_i\dot{\vec{m}} \qquad (5)$$

where $e$ is the electron charge, $\hbar$ is the reduced Plank constant, and $G_r$ is the mixing conductance extracted from Eq. 4. Note that the last term averages to zero on a magnetization precession cycle and thus does not contribute to dc spin-pumping. The corresponding calculated ISHE voltages generated by the pumped spins are shown in Fig. 4A (lines). Theory can quantitatively account for the behavior of the HFM (point (1) in Fig. 4A); however, it fails to explain the SF and the QFM signals. For the upper frequency branch, the theory predicts the same polarization modulation and sign for all ISHE signals arising from coherent spin-pumping, with varying magnitudes for the different points. Figure S6 shows the calculated trajectories of the sublattice magnetizations corresponding to points (1-4) in Fig. 4; sublattice magnetization with overall projection along the applied field displays a larger precession angle than its opposite, resulting in a dynamical net moment against the applied field in all cases (*i.e.*, negative ISHE voltage). Experimentally, the 395GHz SF signal exhibits the expected modulation and sign but is substantially larger than expected from theory. The 240GHz SF has the expected polarization modulation, but not the sign. Finally, as mentioned above, the QFM exhibits neither the modulation nor the sign predicted by theory, confirming that coherent spin-pumping is unlikely to be the mechanism behind the system response after the SF transition.

A possible explanation of the independence of QFM signal on the microwave polarization is that the QFM signal arises from a combined effect of magnetic proximity and thermal spin-current generation. Specifically, it is possible that the ground-state magnetization polarizes the conduction electrons in the Pt so that the majority spins are parallel to the magnetization, hence to the applied magnetic field. At the QFM resonance, microwave heating leads to a temperature gradient in the thickness direction, which in turn generates a spin-polarized current in the Pt that converts into an ISHE voltage.

In the unusual regime of the SF transition, the spin dynamics gradually loses the HFM characteristic while acquiring the QFM behavior. In Fig. 4B, point (2) [point (3)] marks the 395 GHz [240 GHz] SF resonance, where the sign of ISHE follows that of the HFM [QFM] at point (1) [point (4)]. This strongly suggests that there must be a turning point between point (2) and point (3) at which the spin current starts to be dominated by the ground-state magnetization rather than the non-equilibrium spin angular momentum in $MnF_2$. However, the exact location of this critical point and how the eigenmodes evolve in the vicinity of that point can only be determined numerically in the presence of a finite misalignment angle. In Fig. S5, we calculated the net equilibrium magnetization as a function of field, qualitatively verifying the above behavior.

By comparing Fig. 2 and Fig. 4 for the 240 GHz resonance, we further notice that the SF signal [point (3)] is stronger than the QFM signal [point (4)] even though the ground-state magnetization, hence the proximity effect, is apparently smaller at point (3) as the QFM behavior there is not fully developed. A possible reason is that within the narrow window of SF transition, the ground state becomes highly unstable, which appreciably enlarges the dynamical susceptibility $\chi(\omega)$. Under fixed microwave power, the heat production rate is proportional to $[\chi(\omega)]^2$. Therefore, it is natural to expect a significantly larger heating effect at point (3) than at point (4). The subtle behavior in the vicinity of spin-flop transition calls for further systematic measurements with additional microwave frequencies.

**Outlook**

The demonstration of the coherent spin-pumping effect in MnF$_2$/Pt opens the door to advancements in controlling and understanding spin-transfer torques in antiferromagnetic-based systems that may lead to energy-efficient and fault-tolerant spintronic devices operating at THz frequencies. Further exploration of spin-pumping in AF-based systems will enable a thorough understanding of the relation between the structural symmetries of antiferromagnets, the characteristics of their spin dynamics, and the polarization of the associated THz signals, which will help designing the future generation of spintronic applications where antiferromagnets are active players.

**References and Notes:**

1. L. Néel, in *Annales de Physique*. (EDP Sciences, 1936), vol. 11, pp. 232-279.
2. P. Wadley *et al.*, Electrical switching of an antiferromagnet. *Science* **351**, 587-590 (2016).
3. D. Kriegner *et al.*, Multiple-stable anisotropic magnetoresistance memory in antiferromagnetic MnTe. *Nat. Comm.* **7**, 11623 (2016).
4. R. Cheng, D. Xiao, A. Brataas, Terahertz antiferromagnetic spin Hall nano-oscillator. *Phys. Rev. Lett.* **116**, 207603 (2016).
5. R. Khymyn, I. Lisenkov, V. Tiberkevich, B. A. Ivanov, A. Slavin, Antiferromagnetic THz-frequency Josephson-like oscillator driven by spin current. *Sci. Rep.* **7**, 43705 (2017).
6. R. Khymyn, I. Lisenkov, V. S. Tiberkevich, A. N. Slavin, B. A. Ivanov, Transformation of spin current by antiferromagnetic insulators. *Phys. Rev. B* **93**, 224421 (2016).
7. R. Kleiner, Filling the terahertz gap. *Science* **318**, 1254-1255 (2007).
8. J. Nogués, I. K. Schuller, Exchange bias. *Journal of Magnetism and Magnetic Materials* **192**, 203-232 (1999).
9. B. G. Park *et al.*, A spin-valve-like magnetoresistance of an antiferromagnet-based tunnel junction. *Nature materials* **10**, 347 (2011).
10. Y. Wang *et al.*, Room-temperature perpendicular exchange coupling and tunneling anisotropic magnetoresistance in an antiferromagnet-based tunnel junction. *Physical review letters* **109**, 137201 (2012).
11. X. Marti *et al.*, Room-temperature antiferromagnetic memory resistor. *Nature materials* **13**, 367 (2014).
12. I. Fina *et al.*, Anisotropic magnetoresistance in an antiferromagnetic semiconductor. *Nature communications* **5**, 4671 (2014).
13. S. Y. Bodnar *et al.*, Writing and reading antiferromagnetic $Mn_2Au$ by Néel spin-orbit torques and large anisotropic magnetoresistance. *Nature communications* **9**, 348 (2018).
14. M. Meinert, D. Graulich, T. Matalla-Wagner, Electrical switching of antiferromagnetic $Mn_2Au$ and the role of thermal activation. *Physical Review Applied* **9**, 064040 (2018).
15. J. Železný *et al.*, Relativistic Néel-order fields induced by electrical current in antiferromagnets. *Phys. Rev. Letts.* **113**, 157201 (2014).
16. R. Lebrun *et al.*, Tunable long-distance spin transport in a crystalline antiferromagnetic iron oxide. *Nature* **561**, 222-225 (2018).
17. A. S. Núñez, R. Duine, P. Haney, A. MacDonald, Theory of spin torques and giant magnetoresistance in antiferromagnetic metals. *Physical Review B* **73**, 214426 (2006).
18. H. V. Gomonay, V. M. Loktev, Spin transfer and current-induced switching in antiferromagnets. *Physical Review B* **81**, 144427 (2010).
19. R. Cheng, J. Xiao, Q. Niu, A. Brataas, Spin pumping and spin-transfer torques in antiferromagnets. *Physical review letters* **113**, 057601 (2014).

20. P. Ross *et al.*, Antiferromagentic resonance detected by direct current voltages in MnF$_2$/Pt bilayers. *Journal of Applied Physics* **118**, 233907 (2015).
21. C. Kittel, Theory of antiferromagnetic resonance. *Physical Review* **82**, 565 (1951).
22. T. Nagamiya, K. Yosida, R. Kubo, Antiferromagnetism. *Adv. Phys.* **4**, 1 (1955).
23. There is an additional zero-energy mode (spin-superfluid mode) after the SF transition which only becomes visible at low frequencies when the field is applied away from the easy anisotropy axis, but this mode is not accessible within our range of frequencies and will not be further discussed in this work.
24. J. Kotthaus, V. Jaccarino, Antiferromagnetic-Resonance Linewidths in MnF$_2$. *Physical Review Letters* **28**, 1649 (1972).
25. M. Hagiwara *et al.*, A Complete Frequency-Field Chart for the Antiferromagnetic Resonance in MnF$_2$. *International journal of infrared and millimeter waves* **20**, 617-622 (1999).
26. A detailed discussion is presented in the Supplementary Information: S1. Spectroscopy and ISHE Measurements
27. I. S. Jacobs, Spin-Flopping in MnF$_2$ by High Magnetic Fields. *Journal of Applied Physics* **32**, S61 (1961).
28. Y. Tserkovnyak, A. Brataas, G. E. Bauer, Enhanced Gilbert damping in thin ferromagnetic films. *Physical review letters* **88**, 117601 (2002).
29. A. Brataas, Y. Tserkovnyak, G. E. Bauer, B. I. Halperin, Spin battery operated by ferromagnetic resonance. *Physical Review B* **66**, 060404 (2002).
30. Y. Tserkovnyak, A. Brataas, G. E. Bauer, Spin pumping and magnetization dynamics in metallic multilayers. *Physical Review B* **66**, 224403 (2002).
31. E. Saitoh, M. Ueda, H. Miyajima, G. Tatara, Conversion of spin current into charge current at room temperature: Inverse spin-Hall effect. *Applied physics letters* **88**, 182509 (2006).
32. Ø. Johansen, A. Brataas, Spin pumping and inverse spin Hall voltages from dynamical antiferromagnets. *Physical Review B* **95**, 220408 (2017).
33. S. M. Rezende, R. L. Rodríguez-Suárez, A. Azevedo, Theory of the spin Seebeck effect in antiferromagnets. *Phys. Rev. B* **93**, 014425 (2016).
34. S. M. Wu *et al.*, Antiferromagnetic spin Seebeck effect. *Physical review letters* **116**, 097204 (2016).
35. W. Lin, C. Chien, Evidence of pure spin current. *arXiv preprint arXiv:1804.01392*, (2018).
36. The presence of the sample prevents achieving perfect circular polarization, which results on a small residual ISHE signal that varies with the size of the sample.
37. L. Liu, R. Buhrman, D. Ralph, Review and analysis of measurements of the spin Hall effect in platinum. *arXiv preprint arXiv:1111.3702*, (2011).
38. W. Zhang *et al.*, Determination of the Pt spin diffusion length by spin-pumping and spin Hall effect. *Applied physics letters* **103**, 242414 (2013).
39. P. Vaidya, Sub-Terahertz Spin Pumping from an Insulating Antiferromagnet, Version 1.0, Harvard Dataverse (2020); https://doi.org/10.7910/DVN/RFCH1Q

**Acknowledgements:**

We thank Art Ramirez of the Materials Advancement Portal at UC Santa Cruz (https://materials.soe.ucsc.edu/home) for providing us with the MnF$_2$ single crystal. **Funding:** P.V., R.C., D.L., and E.d.B. P.V, etc... acknowledge support by the Air Force Office of Scientific Research under Grant FA9550-19-1-0307. A.B. acknowledges support from the European Research Council via Advanced Grant number 669442 ´Insulatronics´ and the Research Council of Norway, project number 262633 ´QuSpin´. The work at UC Santa Cruz was supported in part by the University of California Multicampus Research Programs and Initiatives grant# MRP-17-454963. A portion of this work was performed at the National High Magnetic Field Laboratory, which is supported by the National Science Foundation Cooperative Agreement No. DMR-1644779 and the State of Florida. **Authors Contributions:** P.V., S.M., J.v.T., D.L. and E.d.B, conceived and designed experiments. P.V., and J.v.T performed high-frequency experiments. S.M. performed materials characterization experiments. S.M. and D.L. provided the samples. P.V. and E.d.B analyzed data and performed numerical simulations. Y. L. and R.C. computed the dynamical response of the system and the corresponding spin pumping. A. B. suggested the experimental detection. Both R.C. and A.B. assisted with the theoretical interpretations of the results. All authors contributed to the discussions and the writing of the manuscript. **Competing interests:** Authors do not have any competing interest. **Data and Materials Availability:** All the data shown in the main text and the supplementary information along with the codes used to analyze the data and generate the fitted plots are available on Dataverse.(*39*)

**Note:** The published version of the paper can be accessed at

https://science.sciencemag.org/content/368/6487/160

**Supplementary materials:** Materials/Methods, Supplementary text, figures and references can be downloaded from

https://science.sciencemag.org/content/suppl/2020/04/08/368.6487.160.DC1

**Figure 1:**

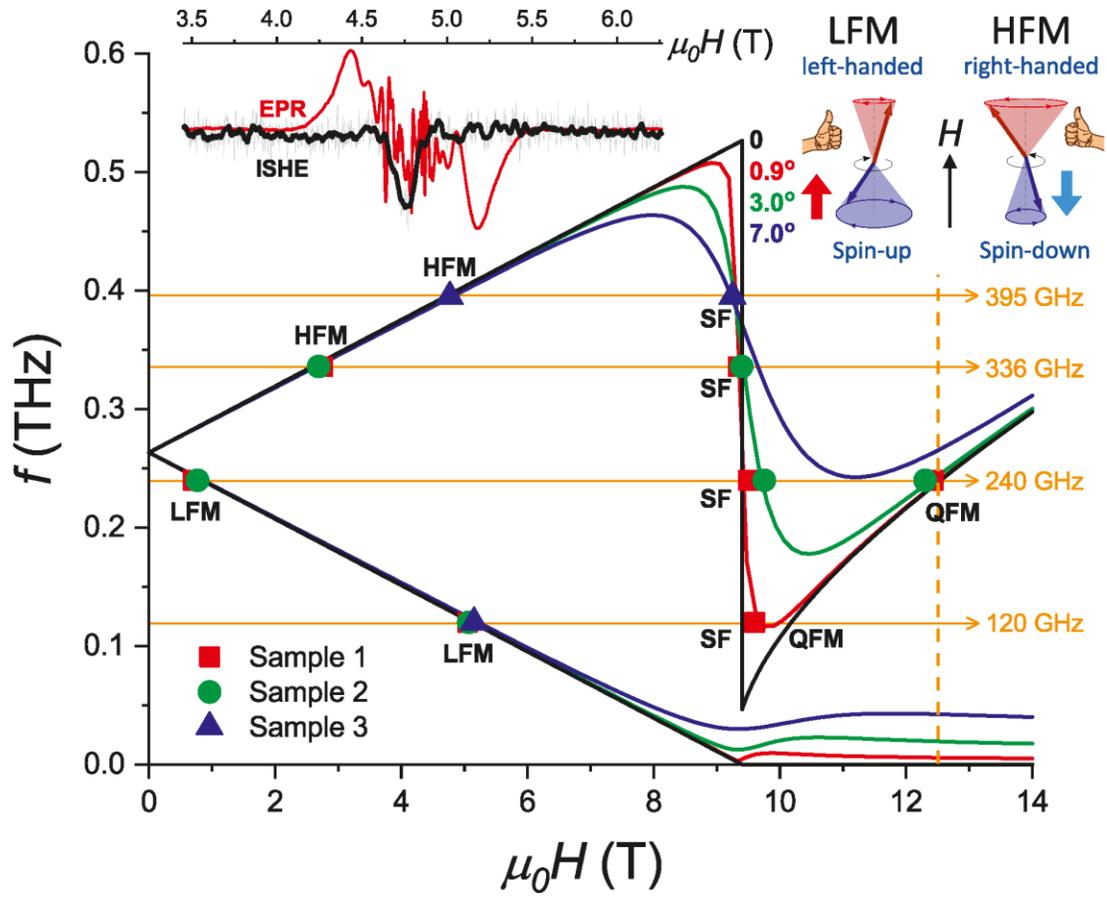

**Figure 2:**

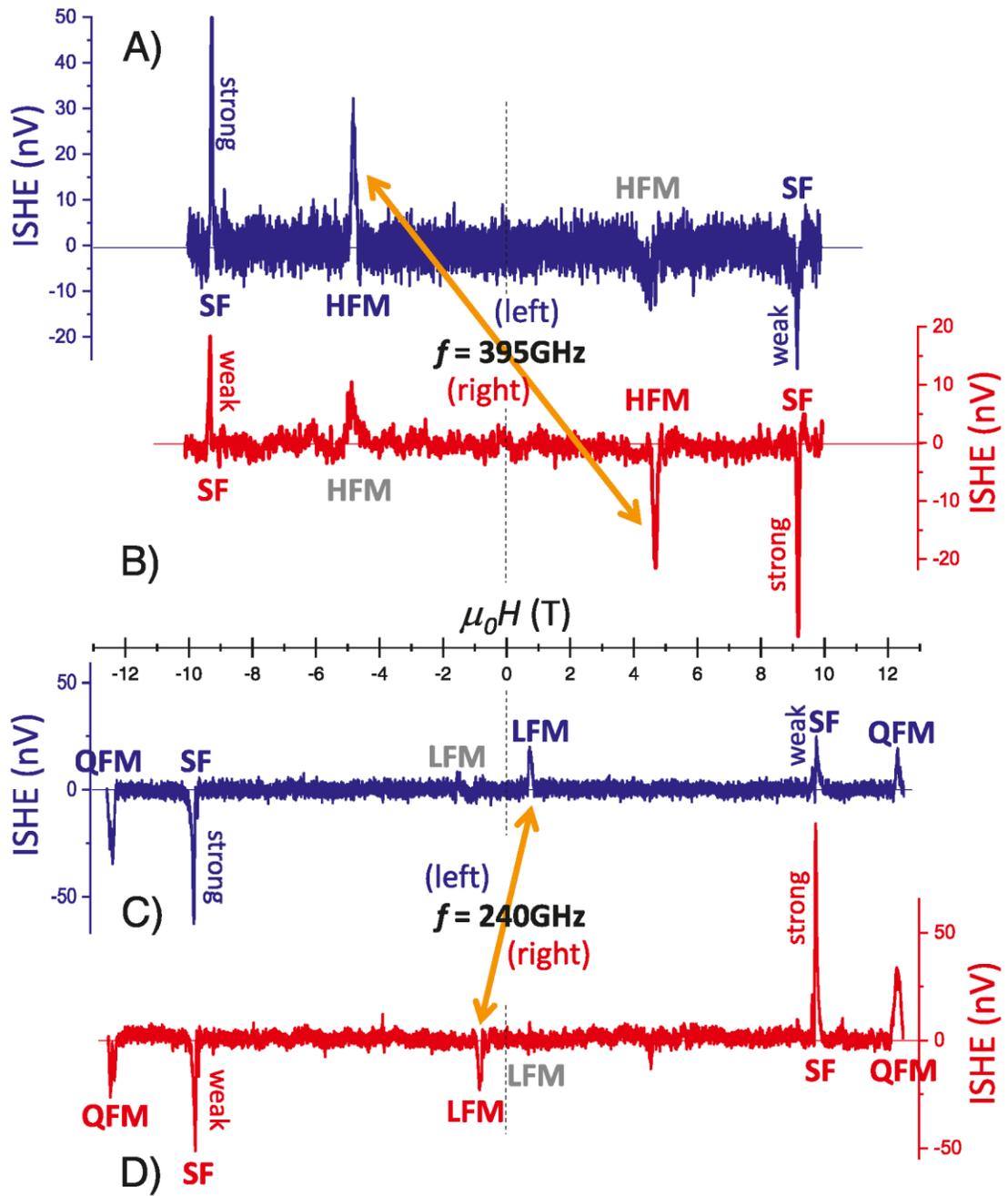

**Figure 3:**

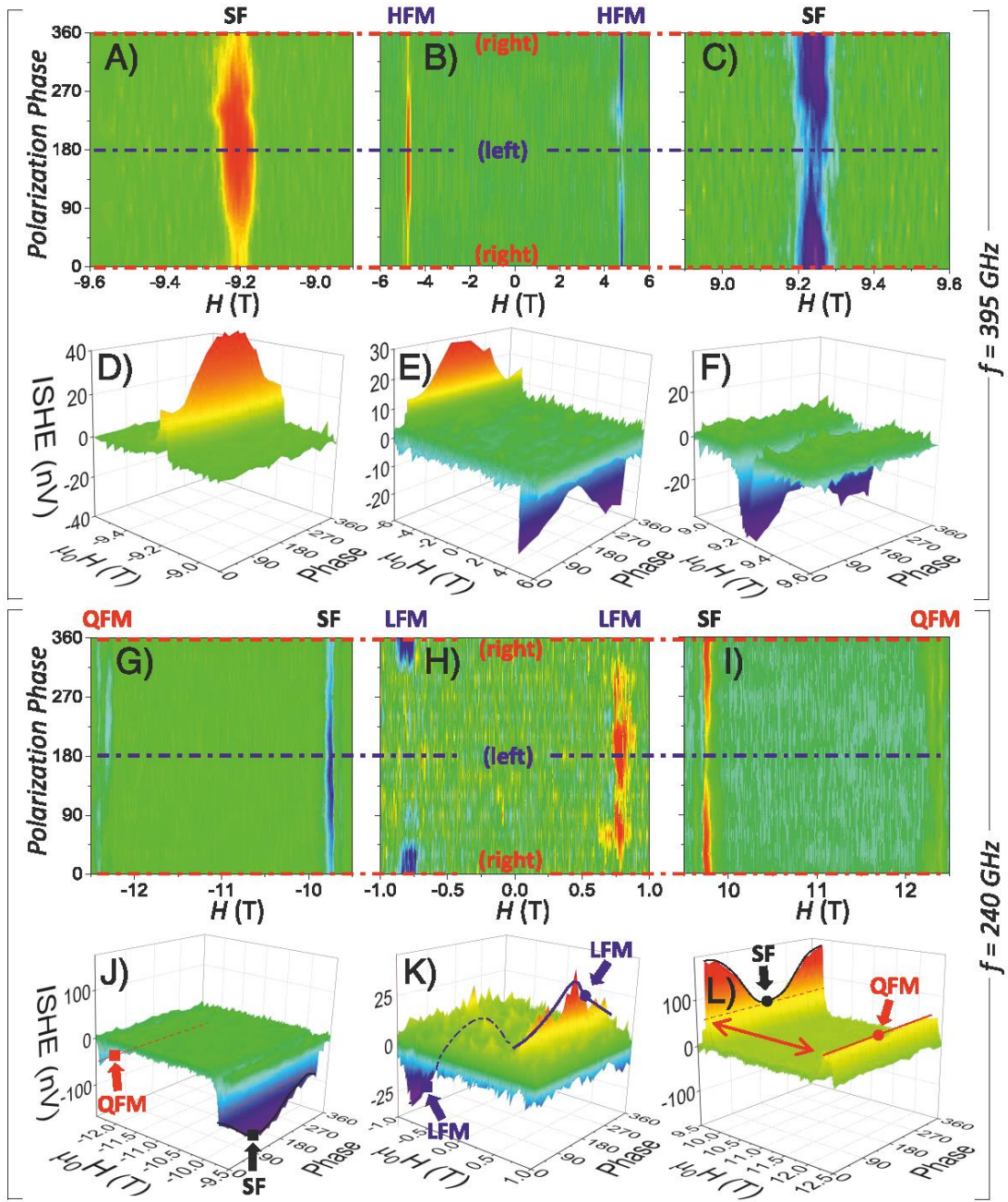

**Figure 4:**

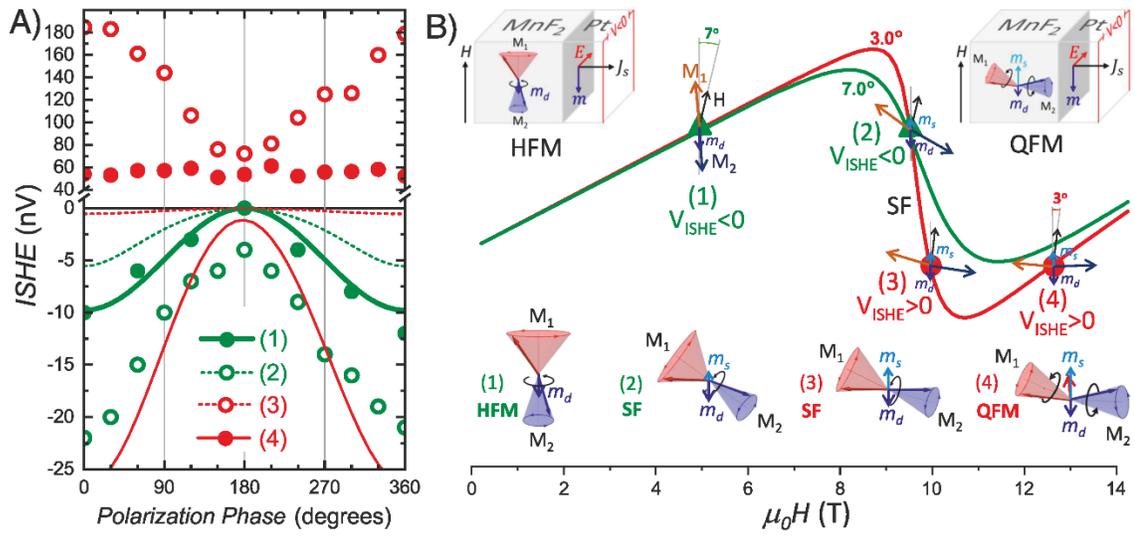

**Figure Legends**:

**Figure 1**: **Antiferromagnetic resonance of MnF₂.** Main panel: Positions of the EPR spectroscopy resonances of MnF₂. The solid curves are the computed resonance frequencies associated with the low- and high-frequency AF modes (sketch in upper right inset), the spin-flop (SF) transition (at $\mu_0 H_{SF} \sim 9.4$ T), and the quasi-ferromagnetic mode (QFM) at high fields. We use the fitting parameters $\mu_0 H_A = 0.82$ T, and $\mu_0 H_E = 47.05$ T in Eq. 1. The different colors correspond to different orientations of the applied magnetic field with respect to the easy anisotropy axis of MnF₂ for each sample. The black curve represents the expected behavior with the field longitudinal to the easy axis. Upper left inset: AFMR spectrum (red) and ISHE response (black) in the adjacent platinum layer corresponding to the high-frequency mode resonance at 395GHz for sample 3 (blue triangle at $\mu_0 H = 4.7$ T).

**Figure 2**: **Inverse Spin Hall Effect in MnF₂/Pt:** ISHE signal obtained in MnF₂/Pt for sample 3 at f = 395 GHz **(A)** left- and **(B)** right-handed circularly polarized microwaves, and for sample 2 at f = 240 GHz microwaves, with both **(C)** left- and **(D)** right-handed circular polarization. A monotonous signal background has been subtracted from all spectra *(26)*. Three distinct features are observed at 240GHz: The low-frequency mode (LFM) at $\mu_0 H = \pm 0.8$ T, the spin-flop (SF) transition resonance at $\mu_0 H = \pm 9.73$ T, and the quasi-ferromagnetic mode resonance at $\mu_0 H = \pm 12.37$ T. Only the high-frequency

mode (HFM) and the SF resonances are observable for 395 GHz at $\mu_0 H = \pm 4.70$ T and $\pm 9.15$ T, respectively, within the available field range.

**Figure 3**: **Circular Polarization Modulation of Spin Pumping:** Evolution of the ISHE signals (magnitude given in the 3D plots) with magnetic field as a function of the polarization of the sub-THz microwaves for **(A-F)** f = 395 GHz and **(G-L)** 240 GHz. Left (right)-handed circular polarization is achieved at 180 (0,360) degrees.

**Figure 4**: **Evolution of Spin Dynamics across the Spin-Flop transition: (A)** The ISHE signals of 395 GHz HFM – point (1), 395 GHz SF – point (2), 240 GHz SF – point (3), and 240 GHz QFM – point (4). Experimental data (dots) and numerical simulation based on coherent spin pumping (curves) agree quantitatively for (1) and qualitatively for (2); (3) and (4) cannot be captured by coherent spin pumping. We used a larger microwave power in the 240 GHz resonances, hence the larger magnitude of the signals for points (3) and (4). **(B)** Illustration of the orientations of the sublattice magnetizations $\vec{M}_1$ and $\vec{M}_2$ and the applied field $H_0$ (with $H_A$ along the vertical z-axis) for four resonances (1-2 for 395 GHz and 3-4 for 240 GHz) representative of the change in AF dynamics in transiting from the HFM into the QFM through the SF region. The upper sketches represent the orientation and spin polarization of the pumped spin current and the induced ISHE electric field with respect to the measuring circuit in the sample. The lower insets illustrate the precessional cones of $\vec{M}_1$ and $\vec{M}_2$ for each of the resonances.